\documentclass[a4paper,10pt,conference]{IEEEtran}
\usepackage[T1]{fontenc} 
\usepackage[utf8]{inputenc} 
\usepackage{graphicx} 
\usepackage{hyperref} 
\usepackage{multirow} 
\usepackage{tabularx} 
\usepackage{color} 
\usepackage{textcomp} 
\usepackage{amsmath} 
\usepackage{amssymb} 
\usepackage{amsfonts} 
\usepackage{amsxtra} 
\usepackage{wasysym} 
\usepackage{isomath} 
\usepackage{mathtools} 
\usepackage{txfonts} 
\usepackage{upgreek} 
\usepackage{enumerate} 
\usepackage{tensor} 
\usepackage{pifont} 
\usepackage{soul} 
\usepackage{booktabs}
\usepackage{cite}
\usepackage{makecell}
\usepackage{threeparttable}
\usepackage[resetlabels,labeled]{multibib}

\newcites{S}{Appendix A. Selected Papers}

\usepackage[colorinlistoftodos,textwidth=10mm,textsize=tiny,obeyFinal]{todonotes}
   \newif\ifdraft
   \draftfalse
   \ifdraft
      \newcommand{\yp}[1]{\todo[color=orange!40]{#1}}
      \newcommand{\mxp}[1]{\todo[color=yellow]{#1}}
   \else
      \newcommand{\yp}[1]{}
      \newcommand{\mxp}[1]{}
   \fi

\begin{document}
\title{Automatic Code Summarization: A Systematic Literature Review}
\author{\IEEEauthorblockN{Yuxiang Zhu, Minxue Pan} 
\IEEEauthorblockA{State Key Laboratory for Novel Software Technology, Nanjing University, China}
\IEEEauthorblockA{zyx@smail.nju.edu.cn, mxp@nju.edu.cn}
}

\maketitle

\begin{abstract}
\textit{Background}: During software maintenance and development, the comprehension of program code is key to success. High-quality comments can help us better understand programs, but they’re often missing or outmoded in today’s programs. Automatic code summarization is proposed to solve these problems. During the last decade, huge progress has been made in this field, but there is a lack of an up-to-date survey. \textit{Aims}: We studied publications concerning code summarization in the field of program comprehension to investigate state-of-the-art approaches. By reading and analyzing relevant articles, we aim at obtaining a comprehensive understanding of the current status of automatic code summarization. \textit{Method}: In this paper, we performed a systematic literature review over the automatic source code summarization field. Furthermore, we synthesized the obtained data and investigated different approaches. \textit{Results}: We successfully collected and analyzed 41 selected studies from the different research communities. We exhaustively investigated and described the data extraction techniques, description generation methods, evaluation methods and relevant artifacts of those works. \textit{Conclusions}: Our systematic review provides an overview of the state of the art, and we also discuss further research directions. By fully elaborating current approaches in the field, our work sheds light on future research directions of program comprehension and comment generation.
\end{abstract}


\section{Introduction}\label{mark-I.}
With the rapid development of scale and complexity of software systems, developers spend about 59\% of their time in program comprehension \cite{xia2017measuring}. However, reading and understanding other peoples' code without good comments can be extremely difficult for software developers. Good program comments can be quite useful in helping developers cooperate or modify others’ code, but for various reasons, high-quality comments are often absent in many software. What’s worse, even if a code snippet is well documented, it also needs to be carefully maintained and renewed.

To address these problems, people propose and design automatic code summarization methods to generate human-readable comments, summaries, commit messages, release notes, etc. These tasks need techniques to abstract high-level actions in code, identify the roles and responsibilities of software units and generate the natural language descriptions.

In this paper, we conducted a systematic literature review in the field of automatic source code summarization using the approach proposed by Kitchenham and Charters (2007) \cite{kitchenham2007guidelines}. A few previous studies of this area have been proposed \cite{nazar2016summarizing,moreno2017automatic}. Our review differs from these studies in the following ways:

\begin{itemize}

\item \textit{Timeframes}: Our review is the most contemporary research \cite{nazar2016summarizing} ering primary studies from January 2010 to January 2019. Nazar et al. \cite{nazar2016summarizing} conducted a review of studies up to April 2016. Moreno and Marcus \cite{moreno2017automatic} didn’t report their timeframe but their selected articles were up to 2016. The difference of timeframe is significant in the field of automatic code summarization, since the first application of artificial neural network to automatic code summarization was published by Iyer et al. in August 2016 [S27]. After Iyer’s work, seven more neural-network-based studies were carried out in the last two years, and they have shown huge potential and efficacy of the deep-learning approach in this area. Nazar’s and Moreno’s survey didn’t capture these neural-network-based studies, which is one of the reasons that motivate us to conduct this review.

\item \textit{Systematic approach}: We followed Kitchenham’s original and strict methods \cite{kitchenham2007guidelines} to conduct this systematic literature review, while Moreno didn’t apply the same process. Nazar et al. defined their research guidelines based on Kitchenham’s approach to conducting their survey. However, we choose to follow the Kitchenham’s path rigorously which has been used widely and proven to be efficient.

\item \textit{Focus}: We focus our attention on text-to-code summarization, which means generate human-oriented summaries from source code artifacts, while Moreno’s and Nazar’s work included text-to-text summarization and code-to-code summarization such as summarizing mailing list and bug reports. What’s more, we concentrated on the technical process of code summarization, while Nazar et al. also paid attention to the applications and tools of their selected studies.

\end{itemize}

The main contributions of this paper are: (i) An analysis of 41 studies focusing on source code summarization from January 2010 to January 2019. Researchers can use these studies to better understand the progress in this field; (ii) A comparison of different approaches of existing researches on summarizing source code; (iii) A synthesis of the current state-of-the-art over source code summarization in our primary studies; (iv) A discussion of the findings in this area and a picture of future research directions.

The rest of the article is organized as follows. First, in Section II we design the process of the study and elaborate our research methodology. In Section III we analyze the extracted data and answer the research questions. Other findings and future research directions are discussed in Section IV. In Section V we summarize and draw the conclusion.

\section{Research Methodology}\label{sec:methodology}

Our systematic literature review was conducted following the guidelines of Kitchenham(2007) \cite{kitchenham2007guidelines}, which structures the activities included in a systematic literature review into three phases: (1) planning, (2) conducting the review, and (3) reporting. The individual tasks performed in each activity are described in the following subsections. Figure 1 shows the approach we employ to carry out this review.

\begin{figure}
\centering
\includegraphics[width=3.4in]{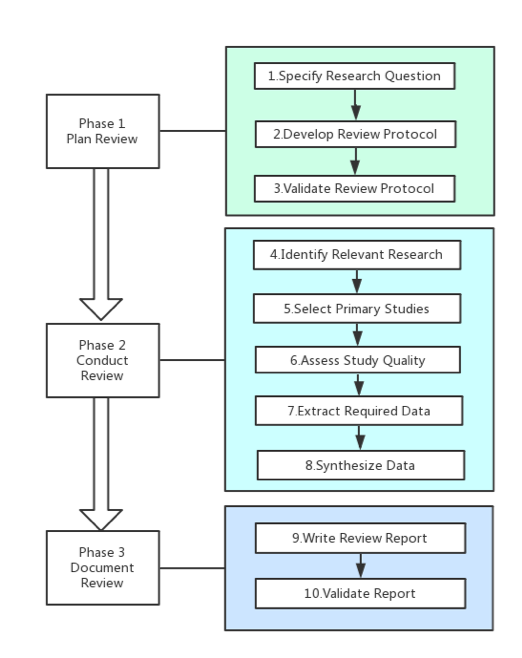}
\caption{Systematic literature review process}
\label{fig:3}
\end{figure}

\subsection{Specify research questions}

Rastkar \cite{rastkar2010summarizing} summarizes that there are two main approaches in generating descriptions of software concerns. The first step is knowledge extraction and deduction, while the second one is the generation of natural language summary. Although software summarization techniques are evolving quickly, those two steps are still suitable for today’s summarization. What’s more, researchers need evaluation techniques to assess the effectiveness of the proposed summarization approaches. Meanwhile, with the development of version control systems, summarizing program change such generating commit messages is an emerging field and also needs prospecting.

To further understand the progresses as mentioned earlier and identify different approaches, the following research questions (RQs) are raised: 

\begin{itemize}

\item RQ1: Which specific techniques are used for extracting information in summarizing source code?

\item RQ2: How to generate natural language descriptions and what kind of summaries can be automatically generated from summarizing source code?

\item RQ3: What evaluating procedures have been used to assess the result in each paper?

\item RQ4: How can we categorize source code artifacts from which we can generate natural language descriptions?

\end{itemize}

\subsection{Development and validation of the review protocol}
The review protocol defines the main activities to conduct the literature review. It consists of four stages: 1. Collecting studies, 2. Selecting primary studies, 3. Analyzing and synthesizing the results, and 4. Reporting the review.

The following digital libraries were searched to collect studies: 

\begin{itemize}

\item ACM digital library,
\item IEEE Explore,
\item ScienceDirect, 
\item Springer Link.

\end{itemize}

We also used Google Scholar to search for more studies. The search terms included \textit{code summarization, summarize code, code mining, automatic document source code, comment generation and summarizing source code change}. The time frame of the search was limited within the last nine years (2010-2019), in which software summarization has grown quickly. We explicitly determined the article inclusion/exclusion criteria in Table I.

\begin{table*}
\caption{Inclusion and Exclusion Criteria}
\begin{tabular}{@{}ll@{}}
\toprule
\multicolumn{1}{c}{\textbf{Inclusion criteria (a paper must be …)}} & \multicolumn{1}{c}{\textbf{Exclusion criteria (a paper cannot be …)}}
\\ \midrule
\makecell*[{{p{8.7cm}}}]{
-	An empirical study; \\
-	Focusing on improving comprehension on source code or other code-related artifacts such as “diff” files of commits; \\
-	Natural language description is the main output (including pseudo code, commit message, release note).
}              & 
\makecell*[{{p{8.7cm}}}]{
-	Focusing on the analysis of natural language artifacts like comments, bug reports, and developer discussions;\\
-	Analyzing source code for purposes other than enhancing human comprehension, such as bug detection, performance analysis;\\
-	Only including analysis of source code without generating descriptions.
}               \\ \bottomrule
\end{tabular}
\end{table*}

\subsection{Identification and selection of primary studies}

Besides the previously described automatic search, we also followed Wohlin’s (2014) snowballing procedure \cite{wohlin2014guidelines} to identify more relevant studies. In the first step, we searched the databases with predetermined keywords as given in the last subsection and then deleted the duplicate records. Our research with the search queries got more than 200 hits, which built up our start set. Then we examined the article title, keywords, abstract and conclusions to filter out irrelevant studies and got 62 articles. We used the snowballing procedure to go through the reference lists of articles in data set to find more relevant ones. We also carried out forward snowballing to identify new papers which cited the paper we have examined, by the tools provided by Google Scholar. After conducting the snowballing procedures, we discovered 33 more relevant papers, so now we had 95 papers altogether. Finally, after using the inclusion/exclusion criteria, 41 studies remained that built up our primary studies. 

\subsection{Extraction of data}

To extract information answering the research questions, we created a data extraction form. First, we classified the articles according to the target software artifacts: source code, program changes, etc. Specifically, we extracted the following data from primary studies:

\begin{itemize}

\item Year of publication;
\item Research method;
\item Preprocessing method;
\item Information extraction method;
\item Summary generation method;
\item Evaluation method;
\item Target software artifacts;
\item Research Conclusion.

\end{itemize}

\subsection{Information analysis and synthesis}

We analyzed 41 selected articles and synthesized the common-used approaches. We recorded the difference and consistency between papers. The result is presented in the next section.

\section{Results}

In the following subsections, we will answer the four research questions separately by analyzing the primary studies.

\subsection{RQ1: Which specific techniques are used for extracting information in summarizing source code?}

The analysis of the 41 studies of summarizing source code identified several specific data extracting techniques used for getting information from source code. The data extracting processes are mainly based on information retrieval, pattern identification, external description mining, natural language process, and machine learning. Table II summarizes our findings of the main distribution of different data extraction techniques. In the following subsections, we will discuss how different methods are used in extracting information.

\begin{table}[!tbh]
\centering
\caption{Data Extraction Method Distribution}
\begin{tabular}{p{3.2cm}p{2.5cm}p{0.2cm}p{0.2cm}}
\toprule
\multicolumn{1}{c}{\textbf{Data Extraction}}   & \multicolumn{1}{c}{\textbf{Paper Reference}} & \multicolumn{1}{c}{\textbf{Freq.}} & \multicolumn{1}{c}{\textbf{Rate}} \\ \midrule
Information Retrieval & {[}S1-8, S10, S21-24, S33, S34, S39, S41{]}  & 17  & 41\% \\ \midrule
Machine Learning and Artificial Neural Network & {[}S8, S25-32, S35-37, S40{]} & 13 & 32\%  \\ \midrule
Stereotype Identification  & {[}S9-15, S38{]}  & 8 & 20\% \\ \midrule
Natural Language Process & {[}S8,,S20-24, S36{]} & 7 & 17\%  \\ \midrule
External Description Usage  & {[}S16-19{]} & 4 & 10\%
\\ \bottomrule
\end{tabular}
\end{table}

\textit{1)	Information Retrieval}

Information Retrieval is widely used to obtain proper information from source code to automatically generate natural language descriptions. In this section, we will discuss different kinds of information retrieval technologies and how they are applied to extract information from source code.

Keyword identification approaches aim to find a keyword list to represent key information of target source code segments. Generally, keyword identification techniques employ text retrieval techniques such as vector space model (VSM) \cite{salton1975vector}, latent Dirichlet allocation(LDA) \cite{blei2003latent}, latent semantic indexing(LSI) \cite{landauer1998introduction} or simply lead terms. Haiduc et al. [S1] used VSM, LSI and lead models to build the keyword list. They evaluated the performance of different techniques and their combinations and found out that the combination of lead and VSM summaries got the highest score. In another article, Haiduc et al. [S2] improved their work, combining text retrieval techniques with structural information in the source code. Rodeghero et al. [S3] further improved the traditional VSM and TF-IDF approach by their eye-tracking study. Using eye-tracking technique, they found out that tested developers spent much more time reading method signatures than method invocations, and spent a bit more time reading method invocation than control flow. So other than basic TF-IDF weight, researchers also weighted the terms based on where they occur, thus achieved a better result than traditional efforts.

Eddy et al. [S4] replicated and expanded the work of Haiduc et al. [S1] They introduced a new topic modeling technique named hierarchical PAM (hPAM) \cite{mimno2007mixtures} for automatic generation of source code summaries. Then they evaluated the quality of their summaries and compared their work with the work by Haiduc [S1], only to find out hPAM is not as good as the VSM methods.

Some works employ simple static analysis techniques \cite{rastkar2010summarizing}, which choose useful information inside classes or methods to generate natural language descriptions. Hammad et al. [S5] extracted information of methods such as the name of local variables and methods invoked. They combined methods’ descriptions to get classes’ description and combined classes’ description to get packages description. Dawood et al. [S34] first converted JAVA program to Abstract Syntax Tree with predefined natural language text Template (AST-W-PDT), and then visited nodes of the tree to obtain information to fill in predefined templates. Kamimura et al. [S6] classified method invocations, compared differences between test cases and identified key facts about the test cases to generate an abstract summary of them.

Topic model can help discover the general topics occurring in the document, which are clusters of similar words. McBurney et al.[S33] represented a software as a call graph and then preprocessed the call graph for the topic model. After that, they used a kind of topic model to deal with the call graph and finally displayed the hierarchical structure of the topics in a web interface. Movshovitz-Attias et al. [S7] used Latent Dirichlet allocation (LDA) and n-gram models for predicting class comments. They creatively exploited code and text tokens in the code to improve the performance of topic model. Fowkes et al. [S8] also used LDA to build a topic model to help automatically fold source code.

\textit{2)	Stereotype Identification}

Stereotypes are abstractions of methods’ or classes’ types and roles in software systems. For example, a method whose responsibility is to construct a class is categorized as constructors.

Moreno et al. [S9] proposed a novel approach called JSummarizer which identified the stereotype of a class by adapting the rules considering the distribution of the methods and their stereotypes in the class. They defined different text templates for different kinds of classes and methods to generate summaries. In another effort [S10] to automatically generate release notes, Moreno et al. reused Jsummarizer [S9] to generate descriptions for newly created classes. 

What’s more, Abid et al. [S11] created a summary template for each method stereotype. Then they extracted main components for the method and identified the method’s stereotype to fill in templates. Li et al. [S12] also identified methods’ stereotypes to help summarize unit test cases. Cortés-Coy et al. [S13] proposed a new method ChangeScribe to detect the changed methods’ responsibility using stereotypes, and generate commit message accordingly.

Some works use micro-patterns to generate descriptions. Micro-patterns are abstractions of method’s function, but a method can be mapped to several micro-patterns. Malhotra et al. [S14] filtered out classes with few dependency relationships and then employed micro-pattern to generate descriptions. Rai et al. [S15] converted source code to XML and identified micro-pattern for methods. They collected method-level information from XML code and used a predefined set of text templates to generate summaries.

\textit{3)	External Description Usage}

External description usage is the way that some researchers use external natural language descriptions and their corresponding source code segments to generate summaries for target source code.

Wong et al. [S16] is the first to use external dataset to facilitate automatic summary generation. They crawled code segments together with their descriptions from a programming Question and Answer site named StackOverflow, which is quite famous among programmers. Then they developed AutoComment to extract code-description mappings and generated descriptions for similar code comment. After that, they developed another approach CloCom [S17], which analyzed existing software repositories. They applied code clone detection techniques to discover similar code segments and used the comments from some code segments to describe the other similar code segments. Then they used NLP to select relevant comment sentences to generate descriptions for the target code segments.

Badihi et al. [S18] used crowdsourcing to collect code-description mappings. They designed a game to motivate people to write summaries for the given code segments. The game also engaged developers to rank others’ summaries. They used NLP techniques to handle those code-description mappings and found the most relevant one to generate summary for the target code segments. Huang et al. [S19] mined version control systems to collect commit-comment pairs. For an input commit, their program will search the database to find a similar commit and fetch the corresponding comment. Then the comment can be used to generate a summary.

\textit{4)	Natural Language Process}

Natural Language Process (NLP) is a technology which can analyze natural language data, identify different components in a sentence and extract structural information from documents. In source code analysis, it’s efficient to use NLP to help understand what does the identifier or the sentence imply.

Sridhara et al. [S20] is the first, by using NLP, to exploit both structural and linguistic clues in the method to summarize the main actions of an arbitrary Java method. Given the signature and body of a method, their automatic comment generator will first preprocess the code segment, by using a novel Software Word Usage Model (SWUM) \cite{hill2009automatically}. SWUM can capture not only the occurrences of words in code, but also their linguistic and structural relationships. Then the generator selected important code segments to generate natural language summaries. They were also the first one to generate comments for Java method parameters automatically [S21]. They found all relevant code segments of an arbitrary parameter. Then they used SWUM to extract information from those appropriate segments. After that, they combined the data to generate comment for the parameter. In another effort, Sridhara et al. [S22] used a novel technique to identify code segments which can learn high-level abstractions of actions and express them as a natural language description. After prepossessing, they used NLP techniques to identify and describe high-level abstract actions of a code sequence. Finally, they generated descriptions from those high-level actions detected.

McBurney et al. [S23] used PageRank algorithm \cite{langville2011google} to discover the most important methods in the given method’s context. They also predefined natural language sentence templates and then used the output of SWUM to fill in these templates.

Wang et al. [S24] utilized an abstract syntax tree(AST) and operations performed on related objects to identify object-related action units in the method. Then they identified the statement that represents the main action of the action unit, which is called the focal statement. After that, they used NLP techniques to convert the focal statement into natural language phrases.

\textit{5)	Machine learning and Artificial Neural Network}

Machine learning can be distinguished into two classes:  supervised learning and unsupervised learning. For a large corpus, crowdsourcing can be used to help label data in supervised summarization. Nazar et al. [S25] first built a suitable corpus of source code segments, and then they invited four students to annotate code segments to extract good summaries. After that, they employed crowdsourcing method to label source code fragment with 21 features. At last, they trained two classifiers: support vector machines (SVM) and Naive Bayes on code fragments to generate summaries with predicted features. Rastkar et al. [S26] identified a set of eight sentence-level features to locate the most relevant sentences to the target change code segments. They further ranked the relevant sentences using an SVM classifier based on the values of these features and then produced a summary by extracting the highest-ranked sentences.

In contrast to supervised learning, unsupervised learning doesn’t need human annotation and can find the knowledge that we don’t know in advance. Fowkes et al. [S8] used topic model, unsupervised learning and natural language process to build a novel autofolding method called TASSAL (Tree-based Autofolding Software Summarization Algorithm), which focuses on optimizing the similarity between the summary and the source code. Autofolding is a technique to automatically create a code summary by folding less informative code regions.

Artificial neural network(ANN) and deep learning are new approaches that emerged recently and has achieved outstanding performance in natural language summarization. Artificial Neural Network has many popular variants, which may achieve a totally different result, like Long short-term memory (LSTM), Convolutional neural networks(CNN) and recurrent neural networks(RNN).

Recurrent neural network is most commonly used in code summarization, and Long short-term memory is a special kind of RNN. Iyer et al. [S27] presented a new model CODE-NN, using LSTM and attention procedure to produce high-level summaries that describe C\# code snippets and SQL queries. The model was trained on a dataset which is crawled from StackOverflow. Their evaluation showed that their model significantly outperformed information retrieval-based studies. Hu et al. [S28] proposed a new approach named DeepCom, which applies Natural Language Processing (NLP) techniques to learn from a massive code corpus and generates comments from learned features. Then they used RNN and LSTM in order to analyze structural information of Java methods for better comments generation. They compared their work with CODE-NN and found out their model outperformed CODE-NN by qualitative analysis. In another effort, Hu et al. [S29] presented a novel RNN-based model TL-CodeSum, which successfully exploited API knowledge together with source code in generating code summarization.

Zheng et al. [S30] used a novel RNN-based attention module called Code Attention to translate code segments to comments, combining domain features of code snippets, such as symbols and identifiers. By focusing on these specific features, Code Attention can understand the structure of code snippets. Liang et al. [S31] made use of RNN to encapsulate the critical structural information of the source code and produce natural language comment. In terms of summarizing program changes, Jiang et al. [S36][S40] has built successful models that can make short commit messages using neural machine translation (NMT).

Convolutional neural network has achieved great success in computer vision field, but is not widely used in code summarization field. Allamanis et al. [S32] introduced a novel convolutional attentional network that successfully performed extreme summarization of source code, where ‘extreme’ means generating extremely concise messages.

\subsection{RQ2: How to generate natural language descriptions and what kind of summary can be automatically generated from summarizing source code?}

Once we can extract or deduct information from source code, we should turn to focus on generating natural language description from known facts. According to Haiduc et al. [S1], a summary can be distinguished into two categories: abstractive and extractive, and these two kinds of summary are often combined to achieve a better description. The Extractive summary consists of unprocessed units such as single word, phrase or code segment. Abstractive information is a kind of high-level, summarized information which isn’t a part of the original document.

We noticed data extracting techniques and summary generating approaches are somehow related. In the following subsections, we’d like to discuss what kind of summary generation technology may be used and what sort of summary will be created.

Table III describes the distribution of natural language description generation methods. We will discuss each of them in the following subsections.

\begin{table}[!tbh]
\centering
\caption{Description Generation Method Distribution}
\begin{tabular}{p{2.7cm}p{3cm}p{0.2cm}p{0.2cm}}
\toprule
\multicolumn{1}{c}{\textbf{Generation Method}}   & \multicolumn{1}{c}{\textbf{Paper Reference}} & \multicolumn{1}{c}{\textbf{Freq.}} & \multicolumn{1}{c}{\textbf{Rate}} \\ \midrule
Template based & {[}S2, S5, S6, S9-15, S18, S20-24, S34, S38, S41{]}  & 19 & 46\% \\ \midrule
Machine learning based & {[}S25-32, S35-37, S40{]} & 12 & 29\% \\ \midrule
Term based & {[}S1-4, S7, S18, S33{]} & 7 & 17\% \\ \midrule
External Description based & {[}S15, S16, S19, S39{]} & 4 & 10\% 
\\ \bottomrule
\end{tabular}
\end{table}

\textit{1)	Term-based Summarization}

Term-based summarization is to generate a summary that contains the most relevant terms for a specific software unit. Most of term-based summarization methods are connected with information retrieval techniques.

Haiduc et al. [S1][S2], Rodeghero et al. [S3] and McBurney et al. [S33] used information retrieval techniques to extract information and then generated keyword list from it in diverse approaches. The generated keyword list captures source code semantics on which developers focus most of the attention.

\textit{2)	Template-based Summarization}

Template-based summarization is the most common natural language summary generation method, and it can use all kinds of extracted intermediate information. In template-based summarization, researchers predefine a set of summary templates and fill in the templates based on the type of the target code segment and other information.

Dawood et al. [S34] and Hammad et al. [S5] just filled in the predefined template with program structural information such as the number of interfaces in a package or what kind of parameter does a method use. Wang et al. [S24] used NLP to identify actions, themes and secondary arguments to fill in templates together with some basic information. McBurney et al. [S23] employed and expanded a Natural Language Generation (NLG) systems proposed by Reiter and Dale \cite{reiter2000building}. NLG is an architecture to translate a set of facts to human-readable natural language sentences.

The Software Word Usage Model (SWUM) is a prevalent model to convert Java method calls into natural language statements \cite{hill2009automatically}. Sridhara et al. [S20][S21][S22], McBurney et al. [S23] and Badihi et al. [S18] have exploited SWUM to create a format-based model which converts source code segments to natural language descriptions. 

Stereotype identification tightly couples with template-based generation. Abid et al. [S11], Moreno et al. [S9][S10], Malhotra et al. [S14] and Rai et al. [S15] identified methods’ or classes’ stereotype based on their property and chose templates accordingly. They filled in the template with other messages such as method signature and return type.

In unit test case summarization, template-based technique is often employed. For example, Li et al. [S12] and Kamimura et al. [S6]  all employed a predefined format to generate a textual summary.

\textit{3)	External-description-based Summarization}

External-description-based summarization uses external data such as comment-code mappings in other repositories or website forums. 

Wong et al. [S16] collected comment-code mappings from StackOverflow and found the most relevant ones to target code segment. Then they filtered and pruned the description from relevant comment-code mappings. At last, they selected the most suitable comment for target code. In another effort, they turned to use comment-code mappings from open source projects in GitHub [S17]. For a target code snippet, they sequentially found most similar code-comment mappings, pruned them, extracted and selected comments. After that, they ranked all matched comments(because the repository database is enormous and there are many similar code segments) and generated the description of target source code. Huang et al. [S19] chose to directly apply the comment of the most similar commit to the input commit.

\textit{4)	Machine-learning-based Summarization}

In early works, supervised learning and unsupervised learning have been used to generate natural language description, but it turns out that machine translation techniques and neural network based natural language generators are more prevalent and efficient. 

As a machine translation model, statistical machine translation (SMT) generates translations by statistical models whose parameters are derived from bilingual text corpora. Oda et al. [S35] expanded SMT to automatically discover the relationship between source code and natural language and used the connection to convert source code into pseudo code. With a source code/pseudo-code parallel corpus, they used the SMT model to train and generate pseudo code from source code.

Neural network was first used by Iyer et al. [S27] in this area. They developed a novel approach named CODE-NN, which exploited an RNN with attention and directly distributed the comments’ words to code tokens. CODE-NN can successfully recommend natural language descriptions using source code snippets collected from StackOverflow. Allamanis et al. [S32] proposed a convolutional neural network to extract translation-invariant features from source code. Unlike common natural language summarization, their approach concentrated on extreme summarization, which means their output summary will be very small (3-terms on average).

Neural machine translation (NMT) is a machine translation model which uses artificial neural network to carry out translation task. In practice, we use sequence-to-sequence (Seq2Seq) model to build language models for both source language and target language. It has achieved great success in machine translation and is now getting increasing attention in source code summarization field. The model consists of an encoder, a decoder, and an attention component, in which the Encoder and Decoder are both RNNs. Hu et al. [S28][S29] and Zheng et al. [S30] employed Seq2Seq Model to translate source code into natural language descriptions. Liang et al. [S31] proposed a new model called Code-GRU, which consists of two RNNs and an intermediate vector and is a kind of variant of Seq2Seq model. It also took structural information into consideration. Jiang et al. [S36] and Loyola et al. [S37] both adopt NMT to translate diff files into short summaries, thus automatically generating commit messages.

\subsection{RQ3: What evaluating procedures have been used to assess the result in each paper?}

How to assess a software summarization process remains the primary challenge in this field. Moreno et al. \cite{moreno2017automatic} have categorized evaluation strategies to different aspects. According to them, evaluation methods can be categorized into online and offline procedures according to whether human is involved in the process. They can also be categorized into extrinsic and intrinsic ones where extrinsic evaluation assesses the impact of the summary on human task performance such as program comprehension and the intrinsic one assesses the inherent quality or performance of summarization output. 

In recent studies, different kinds of evaluation have been combined and used. For different types of summary or different purposes, researchers may use different evaluation methods. In Table IV, each evaluation method and its distribution are displayed. In the subsections below, we will discuss different kinds of evaluation and demonstrate the application of them.

\begin{table}[!tbh]
\centering
\caption{Evaluation Method Distribution}
\begin{tabular}{p{2.5cm}p{3.2cm}p{0.2cm}p{0.2cm}}
\toprule
\multicolumn{1}{c}{\textbf{Evaluation Method}}   & \multicolumn{1}{c}{\textbf{Paper Reference}} & \multicolumn{1}{c}{\textbf{Freq.}} & \multicolumn{1}{c}{\textbf{Rate}} \\ \midrule
Manual Evaluation & {[}S1, S4, S8-10, S12, S13, S15-23, S27, S30, S33-35, S38, S41{]}  & 23 & 56\% \\ \midrule
Statistical Analysis & {[}S8, S18, S22, S24, S25, S27-32, S35-37, S40, S41{]} & 16 & 39\% \\ \midrule
Gold Standard Summary & {[}S2, S3, S6, S8, S10, S24, S39{]} & 7 & 17\% \\ \midrule
Extrinsic Evaluation & {[}S7, S21, S23, S33, S35{]} & 5 & 12\% \\ \midrule
None & {[}S5, S11, S14, S26{]} & 4 & 10\% 
\\ \bottomrule
\end{tabular}
\end{table}

\textit{1)	Manual Evaluation}

Manual Evaluation is the simplest but most commonly used method to assess many properties of summaries. As an intrinsic and online method, manual evaluation is to invite some software practitioners or students to read and rate the summaries in different dimensions and standards. Dimensions include precision, accuracy, compactness, comprehensibility and so on.

Haiduc et al. [S1] employed four students majoring in computer science to evaluate automatically generated summaries. The participants were asked to rate the generated keyword with four levels, from strongly agree to strongly disagree in only a single dimension. Then the researchers took an average score to evaluate the generated summaries. In order to compare results with Haiduc el al. [S1], Eddy et al. [S4] used a similar approach to assess their results.

Sridhara et al. [S20] invited thirteen participants to rate summaries’ accuracy, content adequacy (whether some information is missing) and conciseness (how much unnecessary information is included). In a later work, Sridhara et al. [S21] employed nine human evaluators to judge the generated summaries on accuracy, utility-standalone (whether generated summaries can help understand the role of method parameter), utility-integrated (summaries help understand the intent of the method), necessity.

Similarly, Moreno et al. [S9] asked participants to evaluate content adequacy, conciseness, and understandability. Rai et al. [S15] concentrated on correctness, completeness, non-redundancy, and conciseness, while Dawood et al. [S34] evaluated summaries’ usefulness and closeness to the source code. Wong et al. [S16] concentrated on accuracy, adequacy, conciseness, and usefulness. In a later work of automatic comment generation, Wong et al. [S17] claimed that the current evaluation criteria could be too fine-grained for determining if a comment will be committed and used by developers. They chose to use a more straightforward evaluation approach, which asked participants to rate the generated comments in just three levels: good, fix and bad. Shen et al. [S38] employed human evaluator to compare the performance among primitive commit, other summary and the generated summary of source code change.

It’s interesting that the quantity of levels in rating dimensions is diversified. While most of researchers use a three-level or five-level score, Haiduc et al. [S1] employed a four-level rating, which is called a four-level Likert Scale \cite{likert1932technique}. They abandoned the middle neutral option in order to exclude the situation of central tendency in the answers due to non-committal answers.

\textit{2)	Gold Standard Summary Comparison}

Gold standard summary refers to a human created summary for reference. For a keyword list generated by a trained model, a human-selected keyword list can act as a gold standard summary. Then comparison from the automatic generated summary to gold standard summary measures the effectiveness of the summary generator.

For example, Haiduc et al. [S2] collected six developers’ summaries and used pyramid score \cite{nenkova2004evaluating} to measure similarity among the developers’ summaries and automatically generated summary. Rodeghero et al. [S3] invited human experts to read the Java method and ranked the top five most-relevant keywords from the method. They computed the minimizing Kendall tau distance between the lists and carried out a Mann-Whitney statistical \cite{smucker2007comparison} test for further evaluation. In Wang et al.’s effort [S24], the gold standard summary was not only used to evaluate system-generated descriptions of the source code, but also employed to measure the intermediate result of their generator. Ponzanelli et al. [S39] also used gold standard summary to assess their summaries for various software entities. Kamimura et al. [S6] employed participants to grab key facts in the test case and compared them with automatically grabbed facts.

\textit{3)	Extrinsic Evaluation}

Extrinsic strategies evaluate the influence of the generated summaries when readers perform a particular task. It may directly measure the ability of a summary to improve program comprehension, reading speed, comment typing speed, etc.

Oda et al. [S35] carried out extrinsic research on how their automatically generated summaries help code understanding. They showed unexperienced developers with a function definition and corresponding pseudo-code. Then those participants assigned a 6-level score indicating their impression of how well they understood the code for each sample. Similarly, Mcburney et al. [S23] asked participants to rate for how the summary contains information that helps comprehension.

In contrast, Movshovitz-Attias et al. [S7] used an offline extrinsic evaluation, which successfully minimized its subjectivity. Because their study aimed at reducing programmers’ type-in characters when writing comments, they used a character saving metric so as to quantify the percentage of characters saved by using the model in a word-completion setting, similar to standard code completion tools built inside code editors. Then they used the average percentage of saved characters per comment to assess the model’s efficiency. The result showed that their model successfully saved up to 47.1\% characters.

\textit{4)	Statistical Analysis}

Statistical analysis includes different statistical measures such as precision, recall, F-score, pyramid precision, etc. In most statistical evaluation, those measures of automatically generated summaries are commonly compared with a reference summary, which may be a human-written summary or summary generated by another model. This reference model or system is usually called baseline.

Accuracy, precision, recall, and F-score are simple but efficient ways to quantitatively evaluate the performance of natural language summary generator. Badihi et al. [S18] asked experts to write down lists of keywords for the target method and compared it with the generated keyword list. They used the two sets of lists to calculate accuracy, precision, recall, and F-score and compared these indicators with Rodeghero et al.’s eye-tracking approach [S3], finding out their result is better than eye-tracking approach’s in every index.

Fowkes et al. [S8] asked human annotators to manually fold the source code and used the four index mentioned above to compare their methods and three artificial baselines with the manually annotated result. In Nazar et al.’s supervised learning based study [S25], they used some most common measures by comparing their classifier against random and existing classifier. What’s more, they used precision, recall, and F-score to evaluate the effectiveness of their classifiers.

Besides using precision measure, Sridhara et al. [S22] used specific measures to evaluate their summary generator’s ability to abstract high-level action. They measured the frequency of code fragments in which high-level actions can be identified, claiming that their measured frequency is high enough to demonstrate the broad applicability of their algorithm. They also counted the number of statements that were captured by generated description and the number of phrases in the resulting description, thus calculating the potential reduction in reading detail.

In artificial-neural-network-based studies, machine translation (MT) metrics is the most commonly used evaluation method. Hu et al. [S28][S29], Iyer et al. [S27] and Zheng et al. [S30] used two MT metrics BLEU(Bilingual Evaluation Understudy) score and METEOR (Metric for Evaluation of Translation with Explicit ORdering) to measure the accuracy of automatically generated summaries. Some studies also combined statistical analysis with other evaluation methods. For example, Zheng et al. [S30] also employed a human evaluation to assess the understandability and similarity between generated comments with human-written ones and interpretability.

\subsection{RQ4: How can we categorize source code artifacts from which we can generate natural language descriptions?}

On different levels can developers generate natural language descriptions of code artifact. To answer this research question, we produce a distribution form shown in Table V and list the kinds of target artifacts with a few examples in this section.

\textbf{Variables}: Sridhara et al. [S21] generated parameters’ names and integrated them with method summaries. 

\textbf{Code segments}: For an input code segment, Wong et al. [S16] utilized similar code blocks’ descriptions on StackOverflow. Wang et al. [S24] detected object-related statement sequences and described its high-level action.

\textbf{Methods/Functions}: McBurney et al. [S23] generated natural language summary for a method based on its context. Rai et al. [S15] detected methods’ nano-patterns and generated summaries based on them.

\textbf{Classes}: Moreno et al. [S9] generated summaries for JAVA classes based on their contents, responsibilities, and roles. Malhotra and Chhabra [S14] used micro patterns and dependencies to create summaries.

\textbf{Packages}: Hammad et al. [S5] summarized the services of JAVA packages using template-based techniques. 

\textbf{Test cases}: Li et al. [S12] and Kamimura et al. [S6] used static analysis and code summarization techniques to document unit test cases.

\textbf{Source Code Changes}: Moreno et al. [S10] generated release notes based on different versions of the software program. Cortés-Coy et al. [S13] focused on the generation of commit messages. Buse et al. [S41] proposed a model called DeltaDoc, to describe code modifications by symbolic execution and code summarization.

\begin{table}[!tbh]
\caption{Code Artifact Distribution}
\begin{tabular}{p{2.5cm}p{3.2cm}p{0.2cm}p{0.2cm}}
\toprule
\multicolumn{1}{c}{\textbf{Code Artifact}}   & \multicolumn{1}{c}{\textbf{Paper Reference}} & \multicolumn{1}{c}{\textbf{Freq.}} & \multicolumn{1}{c}{\textbf{Rate}} \\ \midrule
Method & {[}S1-4, S11, S15, S18, S21, S23, S28, S29, S32, S33{]}  & 13 & 32\% \\ \midrule
Code segment & {[}S8, S16, S17, S20, S22, S24, S25, S27, S30, S31, S35, S39{]} & 12 & 29\% \\ \midrule
Code Change & {[}S10, S13, S19, S26, S36-38, S40, S41{]} & 9 & 22\% \\ \midrule
Class & {[}S1, S4, S9, S14, S34{]} & 5 & 12\% \\ \midrule
Test case & {[}S6, S12{]} & 2 & 5\% \\ \midrule
Package & {[}S5{]} & 1 & 2\% \\ \midrule
Variable & {[}S21{]} & 1 & 2\% 
\\ \bottomrule
\end{tabular}
\end{table}

\begin{table*}[]
\caption{List of Primary Studies}
\begin{threeparttable}
\def\arraystretch{1.3}
\begin{tabular}{@{}p{0.1cm}p{3cm}p{3.3cm}p{3.3cm}p{4cm}p{2.33cm}@{}}
\toprule
\textbf{No} & \textbf{Reference} & \textbf{Data Extraction} & \textbf{Description Generation} & \textbf{Evaluation} &
\textbf{Artifact} \tabularnewline
\midrule
1 & Haiduc et al. (2010) & IR (LSI+VSM) & Term based & Manual Evaluation
& Method and class\tabularnewline
2 & Haiduc et al. (2010) & IR (LSI) & Term + Template based & Gold
Standard Summary & Method\tabularnewline
3 & Rodeghero et al. (2014) & IR (VSM + Eye-tracking) & Term based &
Gold Standard Summary & Method\tabularnewline
4 & Eddy et al. (2013) & IR (hPAM) & Term based & Manual Evaluation &
Method and class\tabularnewline
5 & Hammad et al.(2016) & IR (Static Analysis) & Template based & None &
Package\tabularnewline
6 & Kamimura and Murphy (2013) & IR (Static Analysis) & Template based &
Gold Standard Summary, & Test case\tabularnewline
7 & Movshovitz-Attias and Cohen (2013) & IR (LDA + n-gram) & Term based &
Extrinsic Evaluation & N/A\tabularnewline
8 & Fowkes et al.(2017) & IR (Topic Model), Machine learning, NLP & Based on automatically folding code segments & Statistical Analysis, Manual Evaluation, Gold Standard Summary & Code segment\tabularnewline
9 & Moreno et al.(2013) & Stereotype Identification & Template based &
Manual evaluation & Class\tabularnewline
10 & Moreno et al.(2014) & Stereotype Identification and IR (Static
Analysis) & Template based & Gold Standard Summary, and Manual
Evaluation & Program change\tabularnewline
11 & Abid et al. (2015) & Stereotype Identification & Template based &
None & Method\tabularnewline
12 & Li et al. (2016) & Stereotype Identification & Template based &
Manual Evaluation & Test case\tabularnewline
13 & Cortés-Coy et al. (2014) & Stereotype Identification & Template
based & Manual Evaluation & Program change\tabularnewline
14 & Malhotra and Chhabra (2018) & Stereotype Identification & Template
based & None & Class\tabularnewline
15 & Rai et al. (2017) & Stereotype Identification & Template based &
Manual Evaluation & Method\tabularnewline
16 & Wong et al. (2013) & External Description Usage & External
description based & Manual Evaluation & Code segment\tabularnewline
17 & Wong et al. (2015) & External Description Usage & External
description based & Manual Evaluation & Code segment\tabularnewline
18 & Badihi and Heydarnoori (2017) &  External Description Usage  &  Term + template based  &  Manual Evaluation,  and Statistical Analysis & Method 
  \tabularnewline
19 & Huang et al. (2017) & External Description Usage & External
description based & Manual Evaluation & Program change\tabularnewline
20 & Sridhara et al. (2010) & NLP & Template based & Manual Evaluation &
Code segment\tabularnewline
21 & Sridhara et al. (2011) & NLP and IR  &  
Template based  & Extrinsic Evaluation,  and Manual Evaluation & Variable and method  \tabularnewline
22 & Sridhara et al. (2011)  &  NLP and IR  &  Template based  &  Manual Evaluation, and Statistical Analysis  &  Code segment \tabularnewline
23 &  McBurney and McMillan (2014) &  NLP and IR  & Template based (NLG)  &  Extrinsic Evaluation,  and Manual Evaluation &  Method \tabularnewline
24 & Wang et al. (2017) & NLP and IR & Template based & Gold Standard
Summary, and Statistical Analysis & Code segment\tabularnewline
25 & Nazar et al. (2016) & ML (Classifier) & ML based & Statistical
Analysis & Code segment\tabularnewline
26 & Rastkar and G. C. Murphy (2013) & ML (Classifier) & ML based & None
& Program change\tabularnewline
27 & Iyer et al. (2016)  &  ANN (RNN)  &  ML based (RNN)  & Statistical Analysis,  and Manual Evaluation & Code segment \tabularnewline
28 & Hu et al. (2018) & ANN (RNN) & ML based (Seq2Seq) & Statistical
Analysis & Method\tabularnewline
29 & Hu et al. (2018) & ANN (RNN) & ML based (Seq2Seq) & Statistical
Analysis & Method\tabularnewline
30 & Zheng et al. (2017) & ANN (RNN) & ML based (Seq2Seq) & Statistical Analysis, and Manual Evaluation & Code segment \tabularnewline
31 & Liang and Zhu (2018) & ANN (RNN) & ML based (Seq2Seq) & Statistical
Analysis & Code segment\tabularnewline
32 & Allamanis et al. (2016) & ANN (CNN) & ML based (CNN) & Statistical
Analysis & Method\tabularnewline
33 & McBurney et al. (2014) & IR (Topic Model) & Term based & Manual Evaluation, and Extrinsic Evaluation & Method \tabularnewline
34 & Dawood et al. (2017) & IR (Static Analysis) & Template based &
Manual Evaluation & Class\tabularnewline
35 & Oda et al. (2015) & ML(SMT) & ML-based (SMT) & Manual Evaluation, Extrinsic Evaluation, and Statistical Evaluation & Code segment \tabularnewline
36 & Jiang and McMillan (2017) & ML(Classifier), NLP & ML based &
Statistical Evaluation & Program change\tabularnewline
37 & Loyola et al. (2017) & ANN(RNN) & ML based (Seq2Seq) & Statistical
Evaluation & Program change\tabularnewline
38 & Shen et al. (2016) & Stereotype Identification & Template based &
Manual Evaluation & Program change\tabularnewline
39 & Ponzanelli et al. (2015) & IR & External description based & Gold
Standard Summary & Code Segment\tabularnewline
40 & Jiang et al. (2017) & ANN(RNN) & ML based (Seq2Seq) & Statistical
Evaluation & Program change\tabularnewline
41 & Buse and Weimer (2010) & IR & Template based & Statistical
Evaluation and Manual evaluation & Program change\tabularnewline
\bottomrule
\end{tabular}
\begin{tablenotes}
\item[1] IR: Information Retrieval.
\item[2] ML: Machine Learning.
\item[3] Other abbreviations have occurred before.
\end{tablenotes}
\end{threeparttable}
\end{table*}

\section{Discussion}

\subsection{Principle Findings}

Table VI synthesizes the principal findings of our studies. Besides answering the research questions, we also list some other findings below.

\begin{itemize}

\item Source code summarization has been developing very fast in the last decade. In the early 2010s, most of the studies were based on term-based and template-based summarizations. After Iyer et al.(2016) [S27] first employed artificial neural network in this field, many researchers started to use ANN to learn the hidden relationship between structural source code and natural language description, which has been proven surprisingly effective in the automatic code summarization field.
\item Some automatic summary generators rely heavily on the naming quality of identifiers, parameters, methods, and classes. If the source code contains too many local language representations (such as Chinese or Korean transcription), spelling mistakes, or abbreviations, the quality of the generated description may be severely affected. In extensive and finely-supervised software system, this phenomenon may not be an impediment, but it may be a stumbling block in a small-scale system or a legacy system.
\item External data sources in online Q\&A websites, open-source  code repositories and version control systems, etc., have been used widely in this field. Exploiting existing human- written comment of code will significantly benefit the natural language description generation.
\item Over half of the studies use manual evaluation techniques to rate the conciseness, understandability, expressiveness, etc. However, this evaluation techniques is far from objective to assess the ability of natural language generator. The judgment of participants significantly depends on their program ability, experiences, personality and mood. So, offline evaluation and extrinsic evaluation need to be applied in evaluating the summarization. Therefore, the community should build a benchmark or a uniform standard for evaluating natural language summarization.

\end{itemize}

\subsection{Challenges and Future research directions}

Although software summarization techniques have received a lot of attention over the last decade and significant progress has been made in this field so far, future research still needs to be carried out. In the following, we list some possible future research directions which should can be interesting to the community.

\textbf{Neural-network-based and deep-learning-based Study}: Nowadays, although artificial neural network and deep learning have achieved great success in many other fields, structural information is still hard to be properly used in neural-network-based source code summarization. Different from natural language translation, neural machine translation in source code translation should find a way to exploit the advantage of the code structure.

\textbf{Cross-language United Model}: With the development of large-scale software systems, exploiting multiple programming languages is common in a software system or an organization. But if we apply different program summarization techniques for different languages, confusions and inconsistency may arise. Few approaches can handle multiple languages at the same time. Therefore we envision a united model to generate natural language descriptions for various kinds of programming languages.

\textbf{Multi-role-oriented Model}: The information requested by different kinds of readers may be different. For example, authors tend to focus on the responsibility and role description of a method, which can quickly bring their memory back. Software test engineers may need to know the control flow inside the program to employ branch testing, while the method invoker may care about the information regarding how to use the method correctly. How to synthesize different kinds of information remains unanswered.

\textbf{Adaptive Description Generation System}: Comment writing habit and word usage are very personal, which may vary from developer to developer or organization to organization. However, present natural language summarization fails to consider personal typing habit or company’s comment regulation. So, it’s crucial to adapt the system to suit users’ habit. What’s better, an incremental adaptive natural language summarization system for source code may be designed, which will automatically improve its performance after analyzing users’ coding behavior.

\section{Conclusions}

Software summarization is an area which has been growing throughout the last decade. Due to the development of software industry and increase of the software complexity, program comprehension is more and more important, and thus the need for automatic code summarization techniques is growing. 

In this paper, we present the result of a systematic literature review in the field which involves 41 different studies. The result provides useful information for researchers investigating this area to improve code summarization techniques for program comprehension. We hope this review and the proposed future directions can be useful to the community.

\bibliographystyle{IEEEtran}

\nociteS{*}
\bibliographystyleS{IEEEtran}
\bibliographyS{IEEEabrv,selected_papers}

\bibliographystyle{IEEEtran}
\bibliography{IEEEabrv,mylib}

\begin{thebibliography}{10}
\providecommand{\url}[1]{#1}
\csname url@samestyle\endcsname
\providecommand{\newblock}{\relax}
\providecommand{\bibinfo}[2]{#2}
\providecommand{\BIBentrySTDinterwordspacing}{\spaceskip=0pt\relax}
\providecommand{\BIBentryALTinterwordstretchfactor}{4}
\providecommand{\BIBentryALTinterwordspacing}{\spaceskip=\fontdimen2\font plus
\BIBentryALTinterwordstretchfactor\fontdimen3\font minus
  \fontdimen4\font\relax}
\providecommand{\BIBforeignlanguage}[2]{{%
\expandafter\ifx\csname l@#1\endcsname\relax
\typeout{** WARNING: IEEEtran.bst: No hyphenation pattern has been}%
\typeout{** loaded for the language `#1'. Using the pattern for}%
\typeout{** the default language instead.}%
\else
\language=\csname l@#1\endcsname
\fi
#2}}
\providecommand{\BIBdecl}{\relax}
\BIBdecl

\bibitem{haiduc2010use}
S.~Haiduc, J.~Aponte, L.~Moreno, and A.~Marcus, ``On the use of automated text
  summarization techniques for summarizing source code,'' in \emph{2010 17th
  Working Conference on Reverse Engineering}.\hskip 1em plus 0.5em minus
  0.4em\relax IEEE, 2010, pp. 35--44.

\bibitem{haiduc2010supporting}
S.~Haiduc, J.~Aponte, and A.~Marcus, ``Supporting program comprehension with
  source code summarization,'' in \emph{Proceedings of the 32nd ACM/IEEE
  International Conference on Software Engineering-Volume 2}.\hskip 1em plus
  0.5em minus 0.4em\relax ACM, 2010, pp. 223--226.

\bibitem{rodeghero2014improving}
P.~Rodeghero, C.~McMillan, P.~W. McBurney, N.~Bosch, and S.~D'Mello,
  ``Improving automated source code summarization via an eye-tracking study of
  programmers,'' in \emph{Proceedings of the 36th international conference on
  Software engineering}.\hskip 1em plus 0.5em minus 0.4em\relax ACM, 2014, pp.
  390--401.

\bibitem{eddy2013evaluating}
B.~P. Eddy, J.~A. Robinson, N.~A. Kraft, and J.~C. Carver, ``Evaluating source
  code summarization techniques: Replication and expansion,'' in \emph{2013
  21st International Conference on Program Comprehension (ICPC)}.\hskip 1em
  plus 0.5em minus 0.4em\relax IEEE, 2013, pp. 13--22.

\bibitem{article}
M.~Hammad, A.~Abuljadayel, and M.~Khalaf, ``Summarizing services of java
  packages,'' \emph{Lecture Notes on Software Engineering}, vol.~4, pp.
  129--132, 05 2016.

\bibitem{kamimura2013towards}
M.~Kamimura and G.~C. Murphy, ``Towards generating human-oriented summaries of
  unit test cases,'' in \emph{2013 21st International Conference on Program
  Comprehension (ICPC)}.\hskip 1em plus 0.5em minus 0.4em\relax IEEE, 2013, pp.
  215--218.

\bibitem{movshovitz2013natural}
D.~Movshovitz-Attias and W.~W. Cohen, ``Natural language models for predicting
  programming comments,'' in \emph{Proceedings of the 51st Annual Meeting of
  the Association for Computational Linguistics (Volume 2: Short Papers)},
  vol.~2, 2013, pp. 35--40.

\bibitem{fowkes2017autofolding}
J.~Fowkes, P.~Chanthirasegaran, R.~Ranca, M.~Allamanis, M.~Lapata, and
  C.~Sutton, ``Autofolding for source code summarization,'' \emph{IEEE
  Transactions on Software Engineering}, vol.~43, no.~12, pp. 1095--1109, 2017.

\bibitem{moreno2013automatic}
L.~Moreno, J.~Aponte, G.~Sridhara, A.~Marcus, L.~Pollock, and K.~Vijay-Shanker,
  ``Automatic generation of natural language summaries for java classes,'' in
  \emph{2013 21st International Conference on Program Comprehension
  (ICPC)}.\hskip 1em plus 0.5em minus 0.4em\relax IEEE, 2013, pp. 23--32.

\bibitem{moreno2014automatic}
L.~Moreno, G.~Bavota, M.~Di~Penta, R.~Oliveto, A.~Marcus, and G.~Canfora,
  ``Automatic generation of release notes,'' in \emph{Proceedings of the 22nd
  ACM SIGSOFT International Symposium on Foundations of Software
  Engineering}.\hskip 1em plus 0.5em minus 0.4em\relax ACM, 2014, pp. 484--495.

\bibitem{abid2015using}
N.~J. Abid, N.~Dragan, M.~L. Collard, and J.~I. Maletic, ``Using stereotypes in
  the automatic generation of natural language summaries for c++ methods,'' in
  \emph{2015 IEEE International Conference on Software Maintenance and
  Evolution (ICSME)}.\hskip 1em plus 0.5em minus 0.4em\relax IEEE, 2015, pp.
  561--565.

\bibitem{li2016automatically}
B.~Li, C.~Vendome, M.~Linares-V{\'a}squez, D.~Poshyvanyk, and N.~A. Kraft,
  ``Automatically documenting unit test cases,'' in \emph{2016 IEEE
  international conference on software testing, verification and validation
  (ICST)}.\hskip 1em plus 0.5em minus 0.4em\relax IEEE, 2016, pp. 341--352.

\bibitem{cortes2014automatically}
L.~F. Cort{\'e}s-Coy, M.~Linares-V{\'a}squez, J.~Aponte, and D.~Poshyvanyk,
  ``On automatically generating commit messages via summarization of source
  code changes,'' in \emph{2014 IEEE 14th International Working Conference on
  Source Code Analysis and Manipulation}.\hskip 1em plus 0.5em minus
  0.4em\relax IEEE, 2014, pp. 275--284.

\bibitem{malhotra2018class}
M.~Malhotra and J.~K. Chhabra, ``Class level code summarization based on
  dependencies and micro patterns,'' in \emph{2018 Second International
  Conference on Inventive Communication and Computational Technologies
  (ICICCT)}.\hskip 1em plus 0.5em minus 0.4em\relax IEEE, 2018, pp. 1011--1016.

\bibitem{rai2017method}
S.~Rai, T.~Gaikwad, S.~Jain, and A.~Gupta, ``Method level text summarization
  for java code using nano-patterns,'' in \emph{2017 24th Asia-Pacific Software
  Engineering Conference (APSEC)}.\hskip 1em plus 0.5em minus 0.4em\relax IEEE,
  2017, pp. 199--208.

\bibitem{wong2013autocomment}
E.~Wong, J.~Yang, and L.~Tan, ``Autocomment: Mining question and answer sites
  for automatic comment generation,'' in \emph{2013 28th IEEE/ACM International
  Conference on Automated Software Engineering (ASE)}.\hskip 1em plus 0.5em
  minus 0.4em\relax IEEE, 2013, pp. 562--567.

\bibitem{wong2015clocom}
E.~Wong, T.~Liu, and L.~Tan, ``Clocom: Mining existing source code for
  automatic comment generation,'' in \emph{2015 IEEE 22nd International
  Conference on Software Analysis, Evolution, and Reengineering (SANER)}.\hskip
  1em plus 0.5em minus 0.4em\relax IEEE, 2015, pp. 380--389.

\bibitem{badihi2017crowdsummarizer}
S.~Badihi and A.~Heydarnoori, ``Crowdsummarizer: Automated generation of code
  summaries for java programs through crowdsourcing,'' \emph{IEEE Software},
  vol.~34, no.~2, pp. 71--80, 2017.

\bibitem{huang2017mining}
Y.~Huang, Q.~Zheng, X.~Chen, Y.~Xiong, Z.~Liu, and X.~Luo, ``Mining version
  control system for automatically generating commit comment,'' in
  \emph{Proceedings of the 11th ACM/IEEE International Symposium on Empirical
  Software Engineering and Measurement}.\hskip 1em plus 0.5em minus 0.4em\relax
  IEEE Press, 2017, pp. 414--423.

\bibitem{sridhara2010towards}
G.~Sridhara, E.~Hill, D.~Muppaneni, L.~Pollock, and K.~Vijay-Shanker, ``Towards
  automatically generating summary comments for java methods,'' in
  \emph{Proceedings of the IEEE/ACM international conference on Automated
  software engineering}.\hskip 1em plus 0.5em minus 0.4em\relax ACM, 2010, pp.
  43--52.

\bibitem{sridhara2011generating}
G.~Sridhara, L.~Pollock, and K.~Vijay-Shanker, ``Generating parameter comments
  and integrating with method summaries,'' in \emph{2011 IEEE 19th
  International Conference on Program Comprehension}.\hskip 1em plus 0.5em
  minus 0.4em\relax IEEE, 2011, pp. 71--80.

\bibitem{sridhara2011automatically}
------, ``Automatically detecting and describing high level actions within
  methods,'' in \emph{Proceedings of the 33rd International Conference on
  Software Engineering}.\hskip 1em plus 0.5em minus 0.4em\relax ACM, 2011, pp.
  101--110.

\bibitem{mcburney2014automatic}
P.~W. McBurney and C.~McMillan, ``Automatic documentation generation via source
  code summarization of method context,'' in \emph{Proceedings of the 22nd
  International Conference on Program Comprehension}.\hskip 1em plus 0.5em
  minus 0.4em\relax ACM, 2014, pp. 279--290.

\bibitem{wang2017automatically}
X.~Wang, L.~Pollock, and K.~Vijay-Shanker, ``Automatically generating natural
  language descriptions for object-related statement sequences,'' in \emph{2017
  IEEE 24th International Conference on Software Analysis, Evolution and
  Reengineering (SANER)}.\hskip 1em plus 0.5em minus 0.4em\relax IEEE, 2017,
  pp. 205--216.

\bibitem{nazar2016source}
N.~Nazar, H.~Jiang, G.~Gao, T.~Zhang, X.~Li, and Z.~Ren, ``Source code fragment
  summarization with small-scale crowdsourcing based features,''
  \emph{Frontiers of Computer Science}, vol.~10, no.~3, pp. 504--517, 2016.

\bibitem{rastkar2013did}
S.~Rastkar and G.~C. Murphy, ``Why did this code change?'' in \emph{Proceedings
  of the 2013 International Conference on Software Engineering}.\hskip 1em plus
  0.5em minus 0.4em\relax IEEE Press, 2013, pp. 1193--1196.

\bibitem{iyer2016summarizing}
S.~Iyer, I.~Konstas, A.~Cheung, and L.~Zettlemoyer, ``Summarizing source code
  using a neural attention model,'' in \emph{Proceedings of the 54th Annual
  Meeting of the Association for Computational Linguistics (Volume 1: Long
  Papers)}, 2016, pp. 2073--2083.

\bibitem{hu2018deep}
X.~Hu, G.~Li, X.~Xia, D.~Lo, and Z.~Jin, ``Deep code comment generation,'' in
  \emph{Proceedings of the 26th Conference on Program Comprehension}.\hskip 1em
  plus 0.5em minus 0.4em\relax ACM, 2018, pp. 200--210.

\bibitem{hu2018summarizing}
X.~Hu, G.~Li, X.~Xia, D.~Lo, S.~Lu, and Z.~Jin, ``Summarizing source code with
  transferred api knowledge,'' 2018.

\bibitem{zheng2017code}
W.~Zheng, H.-Y. Zhou, M.~Li, and J.~Wu, ``Code attention: Translating code to
  comments by exploiting domain features,'' \emph{arXiv preprint
  arXiv:1709.07642}, 2017.

\bibitem{liang2018automatic}
Y.~Liang and K.~Q. Zhu, ``Automatic generation of text descriptive comments for
  code blocks,'' in \emph{Thirty-Second AAAI Conference on Artificial
  Intelligence}, 2018.

\bibitem{allamanis2016convolutional}
M.~Allamanis, H.~Peng, and C.~Sutton, ``A convolutional attention network for
  extreme summarization of source code,'' in \emph{International Conference on
  Machine Learning}, 2016, pp. 2091--2100.

\bibitem{mcburney2014improving}
P.~W. McBurney, C.~Liu, C.~McMillan, and T.~Weninger, ``Improving topic model
  source code summarization,'' in \emph{Proceedings of the 22nd international
  conference on program comprehension}.\hskip 1em plus 0.5em minus 0.4em\relax
  ACM, 2014, pp. 291--294.

\bibitem{dawood2017source}
K.~A. DAWOOD, K.~Y. SHARIF, and K.~T. WEI, ``Source code analysis extractive
  approach to generate textual summary.'' \emph{Journal of Theoretical \&
  Applied Information Technology}, vol.~95, no.~21, 2017.

\bibitem{oda2015learning}
Y.~Oda, H.~Fudaba, G.~Neubig, H.~Hata, S.~Sakti, T.~Toda, and S.~Nakamura,
  ``Learning to generate pseudo-code from source code using statistical machine
  translation (t),'' in \emph{2015 30th IEEE/ACM International Conference on
  Automated Software Engineering (ASE)}.\hskip 1em plus 0.5em minus 0.4em\relax
  IEEE, 2015, pp. 574--584.

\bibitem{jiang2017towards}
S.~Jiang and C.~McMillan, ``Towards automatic generation of short summaries of
  commits,'' in \emph{Proceedings of the 25th International Conference on
  Program Comprehension}.\hskip 1em plus 0.5em minus 0.4em\relax IEEE Press,
  2017, pp. 320--323.

\bibitem{loyola2017neural}
P.~Loyola, E.~Marrese-Taylor, and Y.~Matsuo, ``A neural architecture for
  generating natural language descriptions from source code changes,''
  \emph{arXiv preprint arXiv:1704.04856}, 2017.

\bibitem{shen2016automatic}
J.~Shen, X.~Sun, B.~Li, H.~Yang, and J.~Hu, ``On automatic summarization of
  what and why information in source code changes,'' in \emph{2016 IEEE 40th
  Annual Computer Software and Applications Conference (COMPSAC)},
  vol.~1.\hskip 1em plus 0.5em minus 0.4em\relax IEEE, 2016, pp. 103--112.

\bibitem{ponzanelli2015summarizing}
L.~Ponzanelli, A.~Mocci, and M.~Lanza, ``Summarizing complex development
  artifacts by mining heterogeneous data,'' in \emph{Proceedings of the 12th
  Working Conference on Mining Software Repositories}.\hskip 1em plus 0.5em
  minus 0.4em\relax IEEE Press, 2015, pp. 401--405.

\bibitem{jiang2017automatically}
S.~Jiang, A.~Armaly, and C.~McMillan, ``Automatically generating commit
  messages from diffs using neural machine translation,'' in \emph{Proceedings
  of the 32nd IEEE/ACM International Conference on Automated Software
  Engineering}.\hskip 1em plus 0.5em minus 0.4em\relax IEEE Press, 2017, pp.
  135--146.

\bibitem{buse2010automatically}
R.~P. Buse and W.~Weimer, ``Automatically documenting program changes.'' in
  \emph{ASE}, vol.~10, 2010, pp. 33--42.

\end{thebibliography}


\begin{thebibliography}{10}
\providecommand{\url}[1]{#1}
\csname url@samestyle\endcsname
\providecommand{\newblock}{\relax}
\providecommand{\bibinfo}[2]{#2}
\providecommand{\BIBentrySTDinterwordspacing}{\spaceskip=0pt\relax}
\providecommand{\BIBentryALTinterwordstretchfactor}{4}
\providecommand{\BIBentryALTinterwordspacing}{\spaceskip=\fontdimen2\font plus
\BIBentryALTinterwordstretchfactor\fontdimen3\font minus
  \fontdimen4\font\relax}
\providecommand{\BIBforeignlanguage}[2]{{%
\expandafter\ifx\csname l@#1\endcsname\relax
\typeout{** WARNING: IEEEtran.bst: No hyphenation pattern has been}%
\typeout{** loaded for the language `#1'. Using the pattern for}%
\typeout{** the default language instead.}%
\else
\language=\csname l@#1\endcsname
\fi
#2}}
\providecommand{\BIBdecl}{\relax}
\BIBdecl

\bibitem{xia2017measuring}
X.~Xia, L.~Bao, D.~Lo, Z.~Xing, A.~E. Hassan, and S.~Li, ``Measuring program
  comprehension: A large-scale field study with professionals,'' \emph{IEEE
  Transactions on Software Engineering}, vol.~44, no.~10, pp. 951--976, 2017.

\bibitem{kitchenham2007guidelines}
B.~Kitchenham and S.~Charters, ``Guidelines for performing systematic
  literature reviews in software engineering,'' 2007.

\bibitem{nazar2016summarizing}
N.~Nazar, Y.~Hu, and H.~Jiang, ``Summarizing software artifacts: A literature
  review,'' \emph{Journal of Computer Science and Technology}, vol.~31, no.~5,
  pp. 883--909, 2016.

\bibitem{moreno2017automatic}
L.~Moreno and A.~Marcus, ``Automatic software summarization: the state of the
  art,'' in \emph{2017 IEEE/ACM 39th International Conference on Software
  Engineering Companion (ICSE-C)}.\hskip 1em plus 0.5em minus 0.4em\relax IEEE,
  2017, pp. 511--512.

\bibitem{rastkar2010summarizing}
S.~Rastkar, ``Summarizing software concerns,'' in \emph{Proceedings of the 32nd
  ACM/IEEE International Conference on Software Engineering-Volume 2}.\hskip
  1em plus 0.5em minus 0.4em\relax ACM, 2010, pp. 527--528.

\bibitem{wohlin2014guidelines}
C.~Wohlin, ``Guidelines for snowballing in systematic literature studies and a
  replication in software engineering,'' in \emph{Proceedings of the 18th
  international conference on evaluation and assessment in software
  engineering}.\hskip 1em plus 0.5em minus 0.4em\relax Citeseer, 2014, p.~38.

\bibitem{salton1975vector}
G.~Salton, A.~Wong, and C.-S. Yang, ``A vector space model for automatic
  indexing,'' \emph{Communications of the ACM}, vol.~18, no.~11, pp. 613--620,
  1975.

\bibitem{blei2003latent}
D.~M. Blei, A.~Y. Ng, and M.~I. Jordan, ``Latent dirichlet allocation,''
  \emph{Journal of machine Learning research}, vol.~3, no. Jan, pp. 993--1022,
  2003.

\bibitem{landauer1998introduction}
T.~K. Landauer, P.~W. Foltz, and D.~Laham, ``An introduction to latent semantic
  analysis,'' \emph{Discourse processes}, vol.~25, no. 2-3, pp. 259--284, 1998.

\bibitem{mimno2007mixtures}
D.~Mimno, W.~Li, and A.~McCallum, ``Mixtures of hierarchical topics with
  pachinko allocation,'' in \emph{Proceedings of the 24th international
  conference on Machine learning}.\hskip 1em plus 0.5em minus 0.4em\relax ACM,
  2007, pp. 633--640.

\bibitem{hill2009automatically}
E.~Hill, L.~Pollock, and K.~Vijay-Shanker, ``Automatically capturing source
  code context of nl-queries for software maintenance and reuse,'' in
  \emph{Proceedings of the 31st International Conference on Software
  Engineering}.\hskip 1em plus 0.5em minus 0.4em\relax IEEE Computer Society,
  2009, pp. 232--242.

\bibitem{langville2011google}
A.~N. Langville and C.~D. Meyer, \emph{Google's PageRank and beyond: The
  science of search engine rankings}.\hskip 1em plus 0.5em minus 0.4em\relax
  Princeton University Press, 2011.

\bibitem{reiter2000building}
E.~Reiter and R.~Dale, \emph{Building natural language generation
  systems}.\hskip 1em plus 0.5em minus 0.4em\relax Cambridge university press,
  2000.

\bibitem{likert1932technique}
R.~Likert, ``A technique for the measurement of attitudes.'' \emph{Archives of
  psychology}, 1932.

\bibitem{nenkova2004evaluating}
A.~Nenkova and R.~Passonneau, ``Evaluating content selection in summarization:
  The pyramid method,'' in \emph{Proceedings of the human language technology
  conference of the north american chapter of the association for computational
  linguistics: Hlt-naacl 2004}, 2004, pp. 145--152.

\bibitem{smucker2007comparison}
M.~D. Smucker, J.~Allan, and B.~Carterette, ``A comparison of statistical
  significance tests for information retrieval evaluation,'' in
  \emph{Proceedings of the sixteenth ACM conference on Conference on
  information and knowledge management}.\hskip 1em plus 0.5em minus 0.4em\relax
  ACM, 2007, pp. 623--632.

\end{thebibliography}
\end{document}